\documentclass[doublecol,figures]{epl2} 
\usepackage{amsmath,amsfonts, amssymb}

\title{Accessibility percolation on n-trees}
\shorttitle{Accessibility percolation on n-trees} 

\author{S. Nowak\inst{1} \and J. Krug\inst{1}}
\shortauthor{S. Nowak and J. Krug}

\institute{                    
  \inst{1} Institute for Theoretical Physics, University of Cologne, D-50937 K\"oln, Germany\\
}

\abstract{Accessibility percolation is a new type of percolation problem inspired by evolutionary biology. 
To each vertex of a graph a random number is assigned and a path through the graph is called accessible if all numbers along the path are in ascending order. For the case when the random variables 
are independent and identically distributed, we derive an
asymptotically exact expression for the probability that there is at least one accessible path from the root
to the leaves in an $n$-tree. This probability tends to 1 (0) 
if the branching number is increased with the height of the tree faster (slower) than linearly. When the random variables are biased such that the mean value increases linearly with the distance from the root of the tree, a percolation threshold emerges at a finite value of the bias.}

\begin{document}

\maketitle

\section{Introduction and outline}
Percolation theory in its modern form was introduced by Broadbent
and Hammersley in 1957 \cite{Broadbent}. Since then percolation 
has become a cornerstone of probability theory and statistical physics, with
applications ranging from molecular dynamics to star formation
\cite{Stauffer_percolation, Sahimi_percolation}, and new variants of the problem 
continue to attract much attention 
(e.g., \cite{Wilkinson1983,Adler1991,Achlioptas2009,Knecht2012}). 
Standard percolation theory is concerned with the loss of global connectivity in a 
graph when vertices or bonds are randomly removed, 
as quantified by  the probability for the existence of an infinite cluster of contiguous vertices.

Here we consider a novel kind of percolation problem inspired by evolutionary biology. Imagine
a population of some lifeform endowed with the same genetic type (genotype). If a mutation occurs,
a new genotype is created which can die out or replace the old one. Provided
natural selection is sufficiently strong, the latter only happens if the new genotype
has larger fitness. As a consequence, on longer timescales
the genotype of the population takes a path through the space of
genotypes along which the fitness is monotonically increasing \cite{Gillespie1984}.
Such a path is called selectively accessible \cite{Weinreich2005,Weinreich_empirical}. 

Since the relationship between genotype and fitness is very complicated and largely unknown, 
it is natural to assign fitness values to genotypes in a random way. Evolutionary accessibility thus
becomes a statistical property of random fitness landscapes \cite{Carneiro,franke_plos,franke_jsp}. 
In recent years it has become possible to experimentally determine fitness landscapes comprising small numbers of genetic loci \cite{Weinreich_empirical,Lunzer_empirical,Poelwijk_empirical}, and the study of evolutionary accessibility is an important tool 
for characterizing and interpreting such data sets \cite{franke_plos,Szendro2013}.

On an abstract level, the problem of interest is defined as follows:
Consider a graph $G$ where a continuous random variable
$w(\sigma)$ is assigned to each vertex $\sigma$. A path of contiguous vertices
$$ \sigma_1 \rightarrow \sigma_2 \rightarrow \ldots \rightarrow \sigma_n $$
through the graph is called accessible if
$$ w(\sigma_1) < w(\sigma_2) < \ldots < w(\sigma_n). $$
Accessibility percolation studies the statistics of such accessible paths, specifically 
the probability for the existence of paths that span the entire graph.
An important difference to standard percolation is that accessibility depends on
the discrete gradient of a globally defined
fitness function $w\colon G\mapsto\mathbb{R}$, rather than on local random
variables \cite{franke_jsp}.
To see this, it is useful to formulate the problem as a kind of bond
percolation on directed graphs, where the edge from $\sigma$ to $\sigma'$
is removed if $w(\sigma) > w(\sigma')$ and a path is called accessible if all edges are present along the path. In the simple case when the $w(\sigma)$ are chosen as independent and identically distributed
(i.i.d.) random variables, this means that a given edge is removed with probability $1/2$, but
in contrast to standard bond percolation the removal of different edges is correlated through
the fitness function. 

In the biological context the natural choice for the underlying graph $G$ is
the $L$-dimensional hypercube, where $L$ is the number of binary genetic loci
\cite{Carneiro,Kauffman_hoc,Gavrilets97,Gokhale09}. The topology of the hypercube is rather peculiar in that the diameter
of the graph (as measured by the length of the longest path) is equal
to its dimension (as measured by the coordination number of a
vertex). Recent numerical and analytic work
has shown that accessibility percolation on the hypercube with i.i.d. random variables is
\textit{critical}, in the sense that minor changes of the model can trigger
a transition from low to high accessibility
\cite{franke_plos,hegarty_martinsson}. 

In order to further explore the role of graph topology in the phenomenology of accessibility percolation,
we here consider the problem on $n$-trees.
The simple structure of the trees allows us to obtain precise analytic bounds
on the probability of existence of accessible paths. Moreover, by scaling the branching number of the 
tree with its height we are able to mimic the structure of the hypercube and place previous 
results \cite{franke_plos,hegarty_martinsson} into a broader context. 

\begin{figure}[t]
\onefigure[width=80mm]{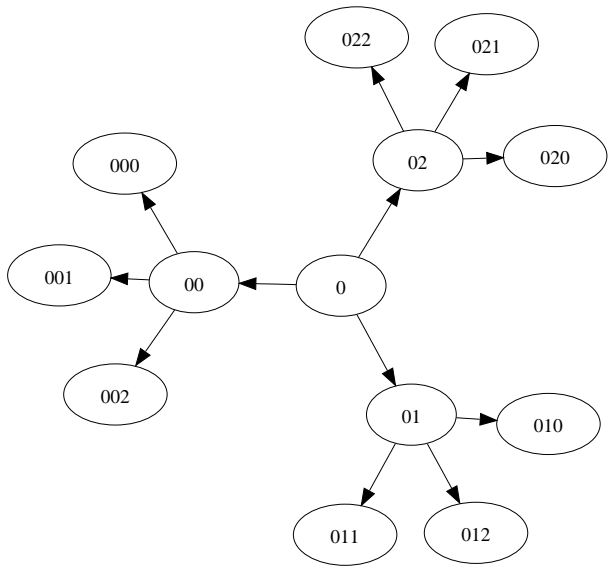}
\caption{An $n$-tree with branching number $n=3$ and height $h=2$.\label{fig:n3L2tree}}
\end{figure}

An $n$-tree or complete $n$-ary tree is a rooted tree where each vertex is connected
to $n$ child vertices, except for the leaves which have no children. The height $h$ is
defined as the distance from the root to the leaves, i.e., any path from the root to
a leaf consists of $h+1$ vertices (cf. fig.~\ref{fig:n3L2tree}).
This graph is very similar to graphs known as Cayley trees or Bethe lattices, though
these are usually defined such that each
vertex has the same coordination number rather than the same number of child vertices, i.e., the root would
have $n+1$ child vertices.

Our results are as follows. First we assume that the random variables $w(\sigma)$ are
drawn independently from a single continuous distribution, a setting known in the evolutionary 
context as the House of Cards (HoC) model \cite{Kauffman_hoc,Kingman_hoc}. As accessibility
is determined solely by the ordering of the random variables on the tree, in this case 
the results are independent of the choice of the underlying probability distribution. 
We will give an exact expression
for the second moment of the number $N{}$ of accessible paths from the
root to the leaves. Furthermore, we will show 
that the probability $\mathbb{P}(N{} \ge 1)$ that there
is at least one accessible path is asymptotically equivalent to $n^h\,/h!$
for large $h$. This result is valid if $n$ is constant as well as if $n=n(h)$
is a function that grows more slowly than linear with $h$. If the growth
is faster than linear, the probability $\mathbb{P}(N{} \ge 1)$ tends
to $1$ for $h \to \infty$. For a linearly growing function $n(h)=\alpha h$ 
there is a percolation threshold, i.e., $\mathbb{P}(N{} \ge 1) \to 0$ for $\alpha < \alpha_\mathrm{c}$
and $\lim_{h \to \infty} \mathbb{P}(N{} \ge 1) > 0$ for $\alpha > \alpha_\mathrm{c}$ with
$\alpha_\mathrm{c} \in [e^{-1}, 1]$.

We then consider the effect of a bias that increases the mean fitness linearly
with the distance from the root, corresponding to the Rough Mount Fuji (RMF) 
model of fitness landscapes in evolutionary biology \cite{franke_plos,Szendro2013,Aita_rmf, Wergen_drift}.
Specifically, we set $w(\sigma) = x_\sigma - \theta d(\sigma)$, where $d(\sigma)$ is the distance
to the closest leaf, $\theta > 0$ is the bias parameter, and the
$x_\sigma$ are i.i.d. random variables
which (for technical reasons to be explained below) are 
drawn from a Gumbel probability density function.
We show that this model displays a percolation threshold at a nonzero
value $\theta_\mathrm{c}$ of the bias, such that
$\lim_{h\to\infty} \mathbb{P}(N \ge 1) = 0$ for $\theta < \theta_\mathrm{c}$
and $\lim_{h\to\infty} \mathbb{P}(N \ge 1) > 0$ for $\theta > \theta_\mathrm{c}$.

\section{Basic properties of n-trees}
Let us recall some basic properties of $n$-trees and their accessibility.
The number of leaves is given by $n^h$.
Since each leaf corresponds uniquely to a path, this is also the number of paths.
In previous work on accessibility percolation on the hypercube, it has usually
been assumed that the value of the destination vertex is the global
fitness maximum \cite{franke_plos,hegarty_martinsson}. Analogously we will assume here
that the starting vertex, i.e., the root, is the global minimum. For the case of i.i.d. random variables 
$w(\sigma)$, there are then $h!$ different equally likely orderings of the values encountered along 
a path, and hence the probability that a given path is accessible is $1/h!$.
By linearity the first moment of the number $N$ of accessible
paths follows immediately \cite{franke_plos},
\begin{align}
 \left< N{} \right> = \frac{n^h}{h!}
\,.
\label{eqn:mean}
\end{align}
Another useful property of $n$-trees is their recursive structure. A tree of height $h+1$ can be thought
of as a root which is connected to $n$ trees of height $h$. This property enables us
to give an explicit but recursive equation for the probability of having no paths from the
root to a leaf. Let us denote the probability that there is no path in a tree of height $h+1$,
given that the root has the value $w(0)=x$, by $Q_h(x)$. A tree of height $h+1$ is not accessible
if each vertex $\sigma$ with distance 1 to the root has either a value $w(\sigma) < x$ or it has
a larger value but the subtree of height $h$ with $\sigma$ as the root is not accessible.
This leads to
\begin{align}
Q_1(x) &= F_0(x)^n\nonumber
\\
 Q_{h+1}(x) &= \left[ F_h(x) + \int_x^\infty f_h(y)\,Q_h(y)\,\mathrm{d}y \right]^n\label{eqn:recursion}
\end{align}
where $f_d(x)$ and $F_d(x)$ are the probability density function and cumulative distribution function, respectively, of
the values $w(\sigma)$ belonging to the vertices which have distance $d$ to the closest leaf. 
For future reference we allow these distributions to depend explicitly on $d$. The probability
of having at least one accessible path in a tree of height $h$ is then given by
\begin{align}
\mathbb{P}(N \ge 1) 
&= 1-\int_{-\infty}^\infty f_h(x)\,Q_h(x)\,\mathrm{d}x 
\,.
\end{align}
So far we did not find a closed analytic solution of the recurrence relation~(\ref{eqn:recursion}).
However, the equation can be solved numerically by approximating the integral by the trapezoidal rule. 
In the following we will use
this to support and supplement our analytic results, 
which are based on the analysis of the moments of $N$.

\section{Calculation of the second moment}

To compute the second moment of $N$ we define indicator
variables $\Theta_i$ for each path $i \in \{1, 2, \ldots, n^h \}$ by
\begin{align}
\Theta_i = \begin{cases}
1, & \text{if the }i\text{-th path is accessible,} \\
0, & \text{else.}
\end{cases}
\end{align}
The $\Theta_i$ are dependent random variables but identically distributed with 
\begin{align*}
 \left< \Theta_i \right> = \left< \Theta_i^2 \right> = \mathbb{P}(\Theta_i = 1) = \frac{1}{h!}
\,.
\end{align*}
This yields $N{} = \sum_{i=1}^{n^h} \Theta_i$ and hence
\begin{align}
 \left< N{}^2 \right>
 = \sum_{i=1}^{n^h} \left< \Theta_i^2 \right> + \sum_{\underset{i \neq j}{i,j=1}}^{n^h} \left< \Theta_i \Theta_j \right>
 = \left< N{} \right> + \sum_{\underset{i \neq j}{i,j=1}}^{n^h} \left< \Theta_i \Theta_j \right>
\,.\label{eqn:secondmoment}
\end{align}
Note that the correlator $\left< \Theta_i \Theta_j \right>$ depends only on the number of vertices which the $i$-th and $j$-th path have in common.
For any two paths $i$ and $j$ (with $i \neq j$) which share $h-k+1$ vertices,
$\left< \Theta_i \Theta_j \right>$ is given by the probability $\pi_k$ that both paths are accessible.
\begin{figure}[thpb]
\centering
\onefigure[width=80mm]{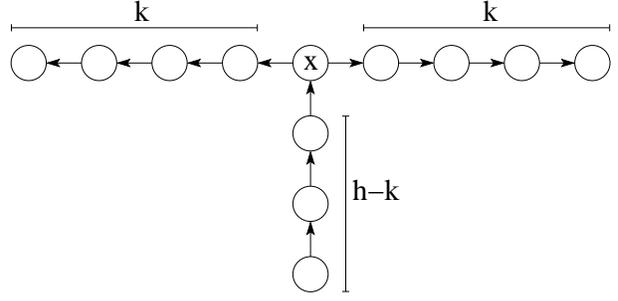}
\caption{The correlation between two paths only depends on the number $h-k+1$ of vertices both paths have in common.\label{fig:twopaths}}
\end{figure}
If the value $w(\sigma)$ of the vertex $\sigma$ where both paths diverge is given by $x$, the two paths
can be decomposed into three independent subpaths (cf. fig.~\ref{fig:twopaths}). All vertices which are
closer to the root than $\sigma$ must have a smaller value than $x$ and all vertices closer to the leaves must have a larger value.
Additionally, all values have to be in ascending order on each subpath. This yields
\begin{align}
&\pi_k = \int_0^1 \frac{x^{h-k-1}}{(h-k-1)!} \left( \frac{(1-x)^{k}}{k!} \right)^2\,\mathrm dx
\label{eqn:pik}
\\
&= \frac{B(h-k, 2k+1)}{(h-k-1)!\,k!^2}
=  {2k \choose k} \frac{1}{(h+k)!}\nonumber
\end{align}
where $B(x,y)$ is Euler's beta function.
To evaluate the sum over the correlators in eqn.~(\ref{eqn:secondmoment})
one also needs the number $m_k$ of pairs of paths that have $h-k+1$ common vertices.
This number can be evaluated with a simple combinatorial consideration:
For the first path, say the left one in fig.~\ref{fig:twopaths}, any leaf can be chosen, i.e., there are $n^h$ possibilities.
The second path shares $h-k+1$ vertices, then there are $n-1$ potential child vertices to choose from
(if one takes the $n$-th vertex, it would also belong to the first path) and finally
one can choose any child vertex until one reaches a second leaf which gives another $n^{k-1}$ possibilities. Altogether there are
$m_k=(n-1)\,n^{h+k-1}$ different pairs. This yields
\begin{align}
&\left< N{}^2 \right> = \left< N{} \right> + \sum_{k=1}^h \pi_k m_k\nonumber\\
&= \left< N{} \right> + \frac{n-1}{n} \sum_{k=1}^h \binom{2k}{k}\frac{n^{h+k}}{(h+k)!}\label{eqn:exact_moment}\\
&\le \left< N{} \right> + \left< N{} \right>^2 + \sum_{k=1}^{h-1} \binom{2k}{k}\frac{n^{h+k}}{(h+k)!}\,.
\label{eqn:correlator}
\end{align}

\section{Probability of having accessible paths}

In order to gain information about the probability for the existence of accessible paths, we will
use the first- and second moment method, i.e., we will apply the following lemma \cite{hegarty_martinsson,Alon_Spencer}:
Let $X$ be a random variable which
takes only integer values $\{ 0, 1, 2, \ldots \}$ and has a finite
second moment. Then the inequality
\begin{align}
 \left< X \right> \ge \mathbb{P}(X \ge 1) \ge \frac{\left< X \right>^2}{\left< X^2 \right>}
\label{eqn:basic_inequality}
\end{align}
holds true. By combining this with eqns.~(\ref{eqn:mean}) and~(\ref{eqn:correlator}) one obtains
the inequality 
\begin{align}
 \left< N{} \right> \ge \mathbb{P}(N{} \ge 1) \ge \frac{\left< N{} \right>^2}{\left< N{} \right> + \left< N{} \right>^2 + S(h)}
\label{eqn:fancy_inequality}
\end{align}
where
\begin{align}
 S(h) &= \sum_{k=1}^{h-1} \binom{2k}{k} \frac{n^{h+k}}{(h+k)!}
\,.
\label{eqn:s_of_h}
\end{align}
We will treat the branching number $n$ as a function of $h$.
Since we are going to distinguish the three cases of linear growth, growth faster
than linear and growth slower than linear, we make the ansatz
\begin{align}
 n = n(h) = h \, \alpha(h) \,.
\end{align}
The three cases correspond to $\alpha(h)=\alpha=\mathrm{const.}$, $\alpha(h) \to \infty$ and
$\alpha(h) \to 0$, respectively. Furthermore, we will assume that $n(h+1) \ge n(h)$.
Using Stirling's formula we obtain for the mean value of $N{}$
\begin{align}
 \left< N{} \right> \approx \frac{n(h)^h}{\sqrt{2 \pi h}\, h^h e^{-h}} = \frac{\left[e\,\alpha(h) \right]^h}{\sqrt{2 \pi h}}
\label{eqn:mean_limit_scaling}
\,.
\end{align}
This expression goes to zero if $\alpha(h) \to 0$ or $\alpha(h)\equiv\alpha \le e^{-1}$
and diverges if $\alpha(h) \to \infty$ or $\alpha(h)=\alpha > e^{-1}$. 

\begin{figure}[thpb]
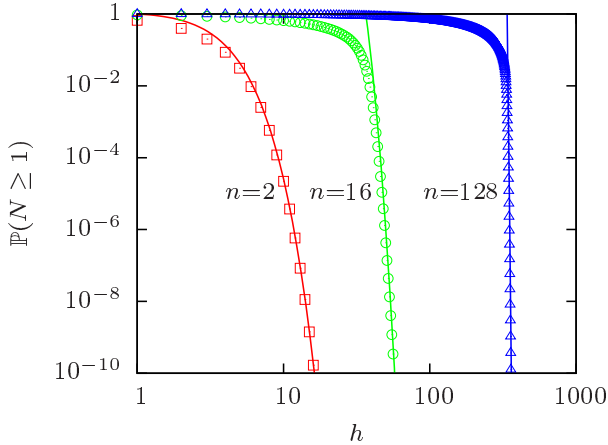

\onefigure[width=80mm]{asymp.epsi}
\caption{The probability $\mathbb{P}(N{} \ge 1)$ as a function of $h$ for constant branching numbers $n$.
Solid lines represent the asymptotic expression $\mathbb{P}(N{} \ge 1) \sim \left< N{} \right>$.
Symbols correspond to the numerical solution of the recurrence relation~(\ref{eqn:recursion}).\label{fig:asymptotics}}
\end{figure}

First we consider the case $\alpha(h) \to 0$ and show that $S(h)$ grows more slowly
than $\left< N{} \right>$.
Stirling's formula applied to eqn.~(\ref{eqn:s_of_h}) leads to
\begin{align*}
S(h) 
&\le \sum_{k=1}^{h-1} \frac{4^k\,{e}^{k+h+1}\,n^{k+h}}{2\pi\,\sqrt{\pi\,k\,(k+h)} \,{\left( k+h\right) }^{k+h}}
\\
&\le 
\frac{n^h}{\sqrt{2\pi (h+1)} (h+1)^h e^{-h}} \sum_{k=1}^{h-1} \left( \frac{4en}{h+1} \right)^k
\\
&\le \left< N{} \right> x \, \frac{1 - x^{h-1}}{1 - x}
\quad\mathrm{with}\quad x = \left( \frac{4en}{h+1} \right)\,.
\end{align*}
Since $x \to 0$ for large $h$ also $S(h) \left< N{} \right>^{-1}$ tends to zero.
This implies according to eqn.~(\ref{eqn:fancy_inequality}) that
$\mathbb{P}(N{} \ge 1)$ is asymptotically equivalent to $\left< N{} \right>$, and both quantities
vanish asymptotically for large $h$ (cf. fig.~\ref{fig:asymptotics}).

The case $\alpha(h) \to \infty$ is a bit more complicated. Similar to the previous case,
we will show that $S(h)$ grows more slowly than $\left< N{} \right>^2$, which implies
that $\mathbb{P}(N{} \ge 1) \to 1$ for $h \to \infty$.
Instead of using Stirling's formula we will estimate
the function $\xi(h)=S(h) \left< N{} \right>_h^{-2}$ recursively:
\begin{align}
&\xi(h+1)= \frac{(h+1)!^2}{n(h+1)^{h+1}} \sum_{k=1}^h \binom{2k}{k} \frac{n(h+1)^k}{(h+k+1)!}\nonumber
\\
&\le \frac{(h+1)^2}{(2h+1)\,n(h+1)} + \frac{(h+1)^2}{(h+2)\,n(h+1)} \xi(h)\nonumber
\\
&\le \frac{1+\xi(h)}{\alpha(h)}\label{eqn:recurrence_xi}
\,.
\end{align}
Since $\alpha(h) \to \infty$, we can fix any constant
$\Omega > 1$ and find $h_0$ such that $\alpha(h) > \Omega$ for all $h > h_0$. Then
it can be shown by induction that
\begin{align*}
 \xi(h) \le \frac{\xi(h_0)}{\Omega^{h-h_0}} + \sum_{k=1}^{h-h_0} \frac{1}{\Omega^k} 
 \xrightarrow{h \to \infty} \frac{1}{\Omega - 1}
\end{align*}
and hence $\xi(h) \to 0$ since $\Omega$ can become arbitrarily large.

Finally we consider the case $\alpha(h)=\alpha=\mathrm{const}$. Using 
the recurrence relation~(\ref{eqn:recurrence_xi}), it can be
shown by induction that $\xi(h)\le (\alpha-1)^{-1}$, given that $\alpha > 1$.
Hence the limit of $\xi(h)$ is finite and according to
eqn.~(\ref{eqn:fancy_inequality}) and~(\ref{eqn:mean_limit_scaling})
we get
\begin{align*}
 \lim_{h \to \infty} \mathbb{P}(N{} \ge 1) \begin{cases}
= 0, & \text{ for } \alpha \le e^{-1}\,,\\
\ge 1 - \alpha^{-1}, & \text{ for } \alpha > 1\,.                                         
\end{cases}
\end{align*}
It is reasonable to assume that the limit is monotonic in $\alpha$. Therefore,
there must be a threshold $\alpha_\mathrm{c}$ in $[e^{-1}, 1]$ such that
$\mathbb{P}(N{} \ge 1) \to 0$ for $\alpha < \alpha_\mathrm{c}$
and $\lim_{h \to \infty} \mathbb{P}(N{} \ge 1) > 0$ for $\alpha > \alpha_\mathrm{c}$.

\section{Effect of a bias}
\label{sec:gumbel}

In this section we assume that the fitness values $w(\sigma)$,
rather than being i.i.d., are distributed
according to $f(x + d(\sigma)\theta)$ where $d(\sigma)$ is the distance from $\sigma$ to the
closest leaf and $f(x)=\exp(-x-\exp(-x))$ is the probability density of the Gumbel distribution.
The reason for this choice is that for this distribution the
probability for the random variables along a path to be in ascending
order (the `ordering probability') is known in closed form
\cite{Wergen_drift}. Using this result it follows that the probability
for a path of length $k$ directed towards the leaves to be accessible is
\begin{align}
P_k(\theta)=a(\theta)^k \, b(\theta, k),
\label{eqn:ordering_probability}
\end{align}
where
\begin{align}
 a(\theta)=1-e^{-\theta}
 \quad\mathrm{and}\quad
 b(\theta, k)=\left[ \prod_{m=1}^{k} (1-e^{-m\,\theta}) \right]^{-1}
\,.
\end{align}
Note that $b(\theta, k)$ is monotonically increasing with $k$ and converges for $k \rightarrow \infty$.
This leads to the bounds
\begin{align}
 1 \le b(\theta, k) \le \underset{k \rightarrow \infty}{\lim} b(\theta, k)
 =: b(\theta)
\label{eqn:boundsforb}
\end{align}
with \cite{Wergen_drift}
\begin{align}
 b(\theta) = \sqrt{\frac{\theta}{2\pi}}\exp\left( \frac{\pi}{6\theta} - \frac{\theta}{24} \right)
\,.
\end{align}
In the following we will show that
$\underset{h \rightarrow \infty}{\lim} \mathbb{P}(N{} \ge 1)=0$ for $a(\theta) < n^{-1}$
and
$\underset{h \rightarrow \infty}{\lim} \mathbb{P}(N{} \ge 1)>0$ for $a(\theta) > n^{-1}$,
and therefore the percolation threshold occurs at 
$\theta_\mathrm{c}=\ln(n)-\ln(n-1)$.

First let $\theta < \theta_c$, i.e., $a(\theta) < n^{-1}$. Then according
to eqn.~(\ref{eqn:basic_inequality}) we get
\begin{align*}
 \mathbb{P}(N{} \ge 1)
 \le \left< N{} \right>
 = P_{h+1}(\theta) n^h
 \le a(\theta) b(\theta) \left[ a(\theta) n \right]^h
 \to 0
\end{align*}
for $h \to \infty$.

\begin{figure}[thpb]
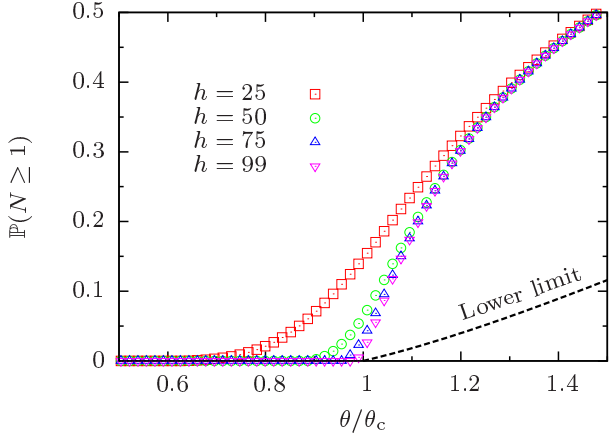

\centering
\onefigure[width=80mm]{transition.epsi}
\caption{Probability of finding accessible paths in a $4$-tree as a
  function of the bias $\theta$. The critical bias is given by
  $\theta_\mathrm{c}=\ln(4/3) \approx 0.288$.
Symbols represent the numerical solution of eqn.~(\ref{eqn:recursion}), the lower bound is given by eqn.~(\ref{eqn:lower_limit_threshold}).
\label{fig:transition}}
\end{figure}

Now let $\theta > \theta_c$. In order to apply eqn.~(\ref{eqn:basic_inequality})
one needs to compute the second moment which in turn requires the calculation
of the probability $\pi_k$ that two paths that share $h-k+1$ vertices are both accessible.
A necessary condition for this is that the $h-k+1$ common vertices are accessible
as well as both separated subpaths which consist of $k$ vertices each (cf. fig.~\ref{fig:twopaths}).
Therefore $\pi_k \le P_{h-k+1}(\theta)\,P_{k}(\theta)^2$ and hence
\begin{align*}
 &\sum_{\underset{i \neq j}{i,j=1}}^{n^h} \left< \Theta_i \Theta_j \right> 
 = \sum_{k=1}^h m_k\,\pi_k
 \le \sum_{k=1}^h m_k P_{h-k+1}(\theta)\,P_{k}(\theta)^2
\\
 &= (n-1) \sum_{k=1}^h n^{h+k-1} a(\theta)^{h+k+1} b(\theta, h-k+1) b(\theta, k)^2
\,.
\end{align*}
After using the upper bound for $b(\theta,k)$ from eqn.~(\ref{eqn:boundsforb}), introducing $y=na(\theta)>1$ 
and evaluating the geometric sum one gets
\begin{align*}
\sum_{\underset{i \neq j}{i,j=1}}^{n^h} \left< \Theta_i \Theta_j \right> 
&\le
\frac{b(\theta)^3}{n} \frac{y^{h+2} (y^h - 1)}{y - 1}
\end{align*}
which finally leads to
\begin{align}
&\mathbb{P}(N{} \ge 1) 
\ge \frac{\left< N{} \right>^2}{\left< N{}^2 \right>} 
= \frac{\left< N{} \right>}{1 + \left< N{} \right>^{-1} \sum_{i \neq j} \left< \Theta_i \Theta_j \right> } \nonumber
\\
&\ge \frac{a(\theta) b(\theta, h+1) y^h}{1 + b(\theta)^2 (y^h-1)\,y\,(y-1)^{-1} } \nonumber
\\
&\xrightarrow{h \to \infty}
\frac{a(\theta)}{b(\theta)}\frac{y-1}{y} > 0
\,.
\label{eqn:lower_limit_threshold}
\end{align}
The lower bound proves that there is a threshold, but
the numerical solution of eqn.~(\ref{eqn:recursion}) indicates that it
is not a good estimate for the exact value of $\mathbb{P}(N{} \ge 1)$
(cf. fig.~\ref{fig:transition}). Nevertheless the transition is
clearly continuous, in qualitative agreement
with the behavior of the bound (\ref{eqn:lower_limit_threshold}). 

Note that the threshold criterion $a(\theta_\mathrm{c}) = n^{-1}$ applies whenever the probability that a
path of length $k$ is accessible can be written as in eqn.~(\ref{eqn:ordering_probability}), where
$b(\theta, k)$ is bounded by two positive functions of $\theta$,
i.e. $c_1(\theta) \ge b(\theta, k) \ge c_2(\theta) > 0$ for all $k$ and $\theta > 0$.
In particular, for $b(\theta, k)=1$ and $a(\theta)=\theta$ it reduces to the case of standard percolation
where vertices are removed with probability $1-\theta$ and a path is accessible if
all vertices are present.
This leads directly to the well-known result $\theta_\mathrm{c} = n^{-1}$ for standard percolation in $n$-trees
\cite{Stauffer_percolation}. A bound on $P_k$ of the form (\ref{eqn:ordering_probability}) has been obtained in 
\cite{hegarty_martinsson} for RMF models with a fairly general class of probability distributions, indicating that
the behavior described here is not restricted to the Gumbel distribution. 

\section{Conclusions}

In this paper we studied evolutionary accessibility as a new type of percolation
problem and considered the statistics of accessible paths on trees.
For the case of i.i.d. random variables, we presented a method for the
exact calculation of the second moment $\left< N^2 \right>$ of the number of accessible paths.
Based on this result, we could show by the second moment method
that the probability $\mathbb{P}(N \ge 1)$ of finding accessible paths
is asymptotically equivalent to the mean value $\left< N \right>$ for large heights $h$
and constant branching number $n$. When $n$ is scaled with $h$, we show that
$\mathbb{P}(N \ge 1)$ goes to 0 or 1 depending on whether $n(h)$ grows more slowly
or faster than linearly.
In case of a linear scaling $n \propto h$ the behavior depends on
the constant of proportionality $\alpha$. We can show that the
limit of $\mathbb{P}(N \ge 1)$ is positive if $\alpha>1$ and zero if $\alpha \le e^{-1}$,
but our method fails for values in between.

The tree with linear scaling $n = \alpha h$ is of particular interest, as it approximates
the topology of the hypercube. There are different ways of relating
the two graphs that lead to different values of the prefactor
$\alpha$. Observing that the coordination number of the hypercube is equal
to its diameter one could set $n = h$ and hence $\alpha=1$. On the other
hand, the total number of pathways traversing a hypercube of size $h$
is $h! \sim (h/e)^h$, which can be equated to the number of paths on
the tree, $n^h$, to yield $\alpha = e^{-1}$. In either case, the comparison
to the tree shows that the hypercube is poised near the threshold
between high and low accessibility. This is consistent with the
rigorous analysis of \cite{hegarty_martinsson}, which shows that the
HoC model on the 
hypercube undergoes an abrupt transition from $\mathbb{P}(N \ge 1)
\approx 1$ to $\mathbb{P}(N \ge 1) \approx 0$ as the fitness value of
the \textit{starting} vertex is increased. 

The different topological structure of hypercubes and trees also
manifests itself in the response to a bias $\theta$ applied to the
random variables along the paths. On the hypercube any
finite bias $\theta > 0$ leads to high accessibility, $\mathbb{P}(N
\ge 1) \approx 1$ \cite{franke_plos,hegarty_martinsson}, while on the tree with fixed branching number $n$
a continuous percolation transition occurs at a threshold value
$\theta_c > 0$. Although the explicit calculation of the threshold was
restricted here to the case of a Gumbel distribution, we expect this
behavior to apply for a large class of distributions (see 
\cite{hegarty_martinsson,Wergen_drift}). 

Our results as well as the inequality~(\ref{eqn:basic_inequality}) indicate that
the probability of having at least one accessible path is, in general, closely related to their mean number.
The left part of the inequality implies that the probability goes to zero
if the mean does. If the mean diverges, a non-vanishing probability
for accessible paths results provided the ratio $\langle N^2
\rangle/\langle N \rangle^2$ remains bounded as the graph size tends
to infinity. This was the case in the examples considered here, but it need not always be true. Explicit
counterexamples are the HoC model on the hypercube without conditions on the
starting vertex, for which $\langle N \rangle = 1$ and  $\mathbb{P}(N
\ge 1) \to 0$ \cite{franke_plos,hegarty_martinsson}, and the block model of protein evolution
\cite{Perelson1995}, for which $\langle N
\rangle \to \infty$ and $\mathbb{P}(N \ge 1) \to 0$ as the hypercube
dimension $L \to \infty$ \cite{Schmiegelt2012}. It is clear from
(\ref{eqn:basic_inequality}) that this behavior requires very large
fluctuations in $N$, and it will be interesting to further explore
these fluctuations and their impact on accessibility. 

Obviously, accessibility percolation can also be studied on other graphs.
But on most structures which are not tree-like it is quite complicated
to calculate the number of paths between two points which is necessary to calculate the second moment
and to apply the inequality~(\ref{eqn:basic_inequality}).
Nevertheless the problem could be analyzed numerically or with other analytical methods.

\acknowledgments
This work was supported by DFG within SPP 1590 and
the Bonn Cologne Graduate School of Physics and
Astronomy. We thank Peter Hegarty and Anders Martinsson
for useful correspondence, and Anton Bovier for discussions.

\end{document}